# Premartensite to martensite transition and its implications on the origin of modulation in Ni$_2$MnGa ferromagnetic shape memory alloy

Sanjay Singh[1], J. Bednarcik[2], S. R. Barman[3], C. Felser[1], and Dhananjai Pandey[4]

[1] Max Planck Institute for Chemical Physics of Solids, Nöthnitzer Strasse 40, D-01187 Dresden, Germany

[2] Photon Sciences, FS-PE, Deutsches Elektronen Synchrotron (DESY), 22607 Hamburg, Germany

[3] UGC-DAE Consortium for Scientific Research, Khandwa Road, Indore, 452001, India.

[4] School of Materials Science and Technology, Indian Institute of Technology ( Banaras Hindu University) , Varanasi-221005, India.

## Abstract

We present here results of temperature dependent high resolution synchrotron x-ray powder diffraction study of sequence of phase transitions in Ni$_2$MnGa. Our results show that the incommensurate martensite phase results from the incommensurate premartensite phase, and not from the austenite phase assumed in the adaptive phase model. The premartensite phase transforms to the martensite phase through a first order phase transition with coexistence of the two phases in a broad temperature interval (~40K), discontinuous change in the unit cell volume as also in the modulation wave vector across the transition temperature and considerable thermal hysteresis in the characteristic transition temperatures. The temperature variation of the modulation wave vector q shows smooth analytic behaviour with no evidence for any devilish plateau corresponding to an intermediate or ground state commensurate lock-in phases. The existence of the incommensurate 7M like modulated structure down to 5K suggests that the incommensurate 7M like modulation is the ground state of $Ni_2MnGa$ and not the Bain distorted tetragonal $L1_0$ phase or any other lock-in phase with a commensurate modulation. These findings can be explained within the framework of the soft phonon model.

**Introduction**

Ni-Mn-Ga ferromagnetic shape memory alloys (FSMA) exhibiting magnetic field induced strains (MFIS) have received considerable attention due to their potential for designing magnetic actuators. [1, 2] *The properties of such alloys as also the crystal structure and the sequence of phase transitions depend very sensitively on the alloy composition*. The stoichiometric $Ni_2MnGa$, despite its brittleness, is the most investigated FSMA as it shows not only extremely large MFIS (~10%) [3, 4] but also large magnetocaloric effect [5] and negative magnetoresistance [6], which are very useful for technological applications. The high temperature phase (austenite) of the stoichiometric $Ni_2MnGa$ is cubic in the space group *Fm-3m* and exhibits a ferromagnetic phase transition at $T_C$ ~ 370 K [7] without any change of crystal structure. On cooling below Tc, the ferromagnetic cubic austenite phase of $Ni_2MnGa$ undergoes premartensitic and martensitic transitions at $T_{PM}$ ~ 260K [8, 9] and $T_M$ ~ 210K [7], respectively, both of which are ferroelastic in nature. On account of Tc being greater than $T_{PM}$ and $T_M$, the stoichiometric alloy exhibits strong magneto-elastic coupling that renormalizes the fourth order term in Landau expansion and makes the two structural transitions first order type with characteristic thermal hysteresis. [1, 7, 10]

Both the premartensite and martensite phases possess modulated structures [8], that is believed to be responsible for the extremely low twinning stresses and hence the large MFIS in $Ni_2MnGa$ [3].

The large MFIS of $Ni_2MnGa$ is intimately linked with the existence of long period modulated structure. Understanding the origin of the modulation is therefore of great significance as it may pave the way for designing new FSMAs. Accordingly, the structure of the modulated premartensite and martensite phases and the origin of modulations in $Ni_2MnGa$ have been topics of intensive research in recent years.[8, 9, 11-18] It is now well established that the modulated structures of the both premartensite and martensite phases are incommensurate in nature. [9, 14, 16] As regards the origin of modulation, there is still controversy. Two different models for the origin of the modulated structures in $Ni_2MnGa$ have been proposed in the literature: an adaptive phase model and a soft-phonon mode based displacive modulation model. In the adaptive phase model, the ground state of the martensite phase is assumed to be $L1_0$ type non-modulated tetragonal structure resulting from Bain distortion (i.e., the lattice deformation strain) of the cubic austenite phase. This stable $L1_0$ phase is believed to undergo periodic nanotwinning in two opposite <1-l0> directions of the austenite {110} planes in order to achieve an optimum austenite – martensite habit plane with minimum interfacial energy. For example, the 7M modulated structure can result from periodic repetition of two twin variants, one consisting of five $L1_0$ type unit cells and the other two $L1_0$ type unit cells.[17] Evidently, the modulated martensite structure in this model corresponds to a kinetically stabilized phase to ensure an optimum habit plane during cubic austenite to non-modulated tetragonal martensite transition. The implicit assumption in this model is that the energy of $L1_0$ twin walls is much less than the austenite – martensite interface energy without the twins. In the soft phonon model, the origin of modulation in the premartensite phase has been related to a *TA2* soft acoustic phonon mode of the austenite phase at q~(1/3 1/3 0). [19, 20-25 ]The soft mode model can account for the ~3M (or equivalently 6M) [18, 26] modulation period of the premartensite phase. [20, 21] A similar soft phonon but in the optical branch leading to the martensite phase with modulated structure has been reported under intense fs laser pulse.[27] The incommensurate nature of modulation of premartensite and martensite phases of $Ni_2MnGa$ [9, 12-16] with non-uniform displacements of various atoms and observation of phason strains [16, 27] seem to favour the soft-phonon mode based mechanism in contrast to the adaptive phase model that is expected to lead to amplitudon strains, uniform displacement of atoms and commensurate modulation only.[17] However, it has been argued that an incommensurate modulation can also result via the adaptive phase model by the formation of discommensurations in the form of stacking faults and antiphase boundaries [17, 18] even as the simulated diffraction patterns involving stacking faults in the 7M commensurate structure are unable to account for the x-ray diffraction (XRD) peak positions correctly. [15]

It is evident from the foregoing that the origin of modulation in the premartensite and martensite phases is an unresolved issue in $Ni_2MnGa$. [15, 16, 17, 18] To settle these unresolved issues related to the origin of modulation, we have carried out temperature dependent high resolution synchrotron x-ray powder diffraction (SXRPD) study of the austenite to premartensite and premartensite to martensite phase transitions. *In an earlier study, Ranjan et al. investigated the structural transitions in near stoichiometric composition $Ni_2Mn_{1.05}Ga_{0.05}$ as a function of temperature but due to the lower resolution of the laboratory based x-ray powder diffractometer, they could not capture the peaks characteristic of the premartensite phase in their x-ray* *diffraction patterns, although their ac susceptibility data shows clear signature of the premartensite phase transition.[11]* The higher resolution SXRPD data used in the present study has enabled us to present unambiguous structural evidence for the coexistence of premartensite and martensite phases in a broad temperature range (~40K) and thermal hysteresis associated with the incommensurate premartensite to the incommensurate martensite phase transition. Our results also unambiguously prove that the martensite phase results from the premartensite phase and not the austenite phase contrary to the basic premise of the adaptive phase model. We also show that at temperatures below the temperature range of phase coexistence, the structure of $Ni_2MnGa$ corresponds to the incommensurately modulated 7M like martensite phase. Further, we show that the temperature dependence of the modulation wave vector of the incommensurate martensite phase shows a smooth analytic behaviour without any "devilish" plateaus corresponding to intermediate lock-in phases down to 5K suggesting that the incommensurate 7M like modulated phase is the ground state of $Ni_2MnGa$ and not the Bain distorted tetragonal $L1_0$ phase postulated in a recent experimental work on adaptive phase model [17] and first principle calculations.[18] Our results clearly disfavour the adaptive phase model of modulation for stoichiometric $Ni_2MnGa$ alloy and support the soft mode model.

**Experimental and Analysis**

Polycrystalline $Ni_2MnGa$ was prepared by standard arc melting technique. The sample composition was checked by EDX, which turns out to be $Ni_{1.99}Mn_{1.01}Ga_{1.00}$. To obtain the characteristic transition temperatures, magnetization as a function of temperature was performed in a low magnetic field of 100 Oe using a superconducting quantum interference device (SQUID) magnetometer. The high resolution synchrotron powder XRD (SXRPD) measurements were performed at a wavelength of 0.20712 Å at P02 beamline in Petra III, Hamburg, Germany. Prior to SXRPD measurements, polycrystalline ingot was ground into fine powder and the powder was further annealed at 773 K under high vacuum of $10^{-7}$ mbar for 10h to remove the residual stresses introduced during grinding. [9, 16, 28, 29] Le Bail and Rietveld analysis of SXRPD data were performed using (3+1) D superspace group approach [30, 31, 32] with JANA2006 software package [33].

**Results and Discussion**

*The low field DC magnetization (M) versus temperature (T) plot of $Ni_2MnGa$ powder shows three anomalies corresponding to the ferromagnetic, premartensite and martensite transitions at Tc ~ 371 K, $T_{PM}= PM_s^c$ ~ 261K and $T_M= M_s^c$ ~ 220K, respectively (Fig. 1) where the subscript "s" stands for the "start temperature" of transition and the superscript "c" stands for "cooling cycle" results.* These premartensite and martensite transition temperatures are consistent with those reported by other workers on stoichiometric $Ni_2MnGa$ [12, 15]. This provides an indirect confirmation that the $Ni_2MnGa$ powder studied here is stoichiometric. *It is evident from Fig.1 that the premartensite start temperature ($PM_s^c$) during cooling (261.4 K) is different from the austenite finish temperature ($A_f^h$) during heating (263.6 K), where the subscript "f" and superscript "h" stand for "finish temperature" and "heating cycle", respectively. The hystereis ($PM_f^h$ - $PM_s^c$) of ~2.5 K (263.6 K -261.4 K) confirms the first order nature of this transition. However, such a small hysteresis also indicates that it is a weakly first order transition. It is intriguing to note that the M(T) curves during heating and cooling do not coincide above $PM_s^c$. A similar magnetization behaviour has also been reported earlier for bulk $Ni_2MnGa$[Ref.34]. In* *our synchrotron XRD measurements, to be discussed below, we could not see any signature of premartensite phase present above $PM_s$ during heating or cooling, which rules out the possibility of long ranged ordered premartensite phase coexisting with the austenite phase being responsible for the difference between M(T) plots during heating and cooling above $PM_s^c$. NMR studies have, however, revealed local correlations. This phenomenon requires further investigation but is outside the scope of the present study.*

Evolution of the high resolution SXRPD patterns of stoichiometric $Ni_2MnGa$ recorded at various temperatures in the range 300K-108K is shown in Fig.2 for two selected 2θ ranges. The room temperature (298K) SXRPD pattern does not reveal any splitting of the XRD peaks as expected for the cubic austenite phase of $Ni_2MnGa$. The excellent fit between the observed and calculated profiles, as obtained by Rietveld refinement of the room temperature austenite phase, confirms the cubic structure in the Fm-3m space group (see Fig 3a). The cell parameter (a= 5.82445(1) Å) obtained by us is in good agreement with those reported by earlier workers. [7, 11] On cooling the sample below the premartensite phase transition temperature ($T_{PM}$= 261 K), several smaller intensity peaks appear (e.g. two such peaks are marked in the inset of Fig. 2(a) and Fig 2(b)), while the austenite cubic peaks remain nearly unaffected. The low intensity peaks are the satellites that appear due to the modulated nature of the premartensite phase of $Ni_2MnGa$ [9]. On lowering the temperature below 248K, the intensity of the peaks charactersitic of the premartensite phase increases indicating that the premartensite phase fraction is growing. There is no evidence for structural transformation to the martensite phase up to 228 K, as all the peaks up to this temperature could be identified with the premartensite phase. Since the structure of the premartensite phase is now known to be incommensurate with 3M like modulation [9, 13, 15], we carried out Rietveld refinement of the modulated structure of the premartensite phase at 228K using the (3+1) D superspace group approach taking into account both the main and the satellite reflections for the superspace group *Immm* (00γ) s00. The excellent fit between the observed and calculated intensities shown in Fig. 3b confirms the incommensurate nature of modulation. The modulation wave vector (**q**) obtained after the refinement is found to be **q**= 0.33769(10)**c***= (1/3+δ) **c***, where δ= 0.00435 is the degree of incommensuration of the modulation of the premartensite phase at 228K. The refined lattice parameters at 228K are : a=4.11455(6) Å, b= 5.81843(9) Å and c= 4.11340(8) Å. The refined lattice parameters and modulation vector are in very good agreement with our earlier results.[9]

At ~218 K, i.e. just below the martensite start temperature $M_s^c$ (= 220 K), additional peaks start appearing (marked with symbol M in Fig.2), which coexist with the premartensite peaks (marked with symbol P). These new peaks are due to the martensite phase whose intensity grows at the expense of the premartensite phase peaks up to 168 K below which the premartensite phase peaks disappear completely. The martensite structure is stable up to the lowest temperature (108 K) at which the data was collected, as no additional reflections appear. Fig.3c depicts the results of (3+1) D superspace Rietveld refinement using *Immm* (00γ) s00 superspace group for the 108 K data. The unit cell parameters obtained after refinement (a= 4.21777 (3) Å, b= 5.54848 (5) Å and c= 4.18741 (3) Å) are in good agreement with the previously reported values [16]. The refined value of the incommensurate modulation vector at 108K was found to be **q**= 0.43083 (8) **c***= (3/7+δ) **c***, where δ= 0.00225 is the degree of incommensuration of the martensite structure at 108 K. This value is also close to the value reported in the literature[17]. Thus the modulated structure of $Ni_2MnGa$ remains incommensurate (7M like) below 168K.

Having discussed the incommensurate nature of the modulated structures of the premartensite and martensite phases, we now proceed to consider the structure of $Ni_2MnGa$ in the temperature range 178K to 218K over which there are many more peaks than anticipated for the pure premartensite or martensite structures. We consider the 218 K data as a representative of this temperature range and present the fits between calculated and observed intensities for (3+1) D superspace Rietveld refinement considering five different structural models. First we considered pure 3M like incommensurate premartensite structure model. This model misses out the new peaks, as evident from Fig 4(a) (see also the insets), and can therefore be rejected. A similar refinement carried out assuming pure 7M like incommensurate modulated structure also could not account for all the peaks including some of the more intense peaks (see Fig 4(b)). As a next step, we considered coexistence of martensite and cubic austenite structures proposed by Ranjan et al.[11] in *near stochiometric composition $Ni_2Mn_{1.05}Ga_{0.95}$* for the temperature range under discussion. *The x-ray diffraction data used by Ranjan et al. for near stoichiometric composition were collected using low resolution laboratory source based x-ray diffraction machine which did not reveal the premartensite peaks. As a result, they, like Brown et al (**Ref 8**) who used medium resolution neutron diffraction data, could not capture the subtle features of the incommensurate nature of modulations discussed in **Ref.16**. Accordingly, Ranjan et al used commensurate model of Brown et al (**Ref 8**) for the modulated structure of the martensite phase. Since the structure of the martensite phase is now settled as incommensurate 7M like using high resolution synchrotronn XRD data[16], we considered coexisting cubic austenite and incommensurate 7M like martensite structures and carried out refinements, instead of coexisting commensurate 7M modulated martensite phase with the cubic austenite phase considered by Ranjan et al that is* *incorrect. The results shown in Fig. 4(c) clearly disfavour the austenite and martensite coexistence model of Ranjan et al even after taking into account the incommensurate nature of modulation in the martensite phase.* Finally, we consider the coexistence of the incommensurate 3M like premartensite and incommensurate 7M like martensite structures in our refinement. This model not only gives an excellent fit between the observed and calculated profiles accounting for all the peaks (see Fig 4(d)) but also gives the best goodness of fit (*S*). Since *Ranjan et al were unable to resolve the premartensite phase peaks, they could not capture the coexistence of premartensite and martensite phases discussed in our present work.* We thus conclude that the 3M like premartensite and 7M like martensite phases, both with incommensurate modulations, coexist at 218K. Using superspace Rietveld refinement, the coexistence of 3M like and 7M like incommensurate structures was confirmed in the entire temperature range from 218 to 178 K. *Phase coexistence across a structural phase transition temperature is a typical characteristic of a first order phase transition where the high temperature phase can coexist metastably well below the thermodynamic phase transition temperature as a supercooled phase during cooling due to nucleation kinetics. In a similar way, the low temperature phase can coexist metastably as a superheated phase.*

In the classical literature on martensites, the temperature at which the martensite phase first appears is called martensite start temperature Ms ($M_s^c$). Below $M_s^c$, the austenite and martensite phases coexist until it attains the martensite finish temperature $M_f$ ($M_f^c$) below which it is purely martensite phase. On heating, the temperature at which the austenite phase first appears is called austenite start temperature $A_s$. The martensite phase fraction decreases on heating above $A_s$ while austenite phase fraction increases until the austenite finish temperature $A_f$ above which only the austenite phase persists. The situation in Ni$_2$MnGa is quite different from the classical martensites as the martensite phase results from a premartensite phase and not the austenite phase. The variation of phase fraction of the premartensite and martensite phases, as obtained by superspace Rietveld refinement, with temperature is shown in Fig 6(a) for both cooling and heating cycles. It is evident from this figure that on cooling, the martensite appears around M$_s^c$ ~218K in coexistence with the premartensite phase. With further decrease in temperature, the premartensite phase fraction gradually decreases and approaches zero around M$_f^c$ ~168K. Similarly, on heating, the premartensite phase appears at PM$_f^h$ ~ 218K and the complete transformation to premartensite phase occurs at PM$_f^h$ ~258K (as determined from the thermal evolution of XRD patterns shown in Fig 1). The fact that M$_f^c$ ≠ PM$_s^c$ and PM$_f^h$ ≠ M$_s^c$ clearly shows the hysteresis in the start and finish temperatures during cooling and heating cycles as expected for a first order phase transition.

A first order phase transition is also accompanied with a discontinuous change in the unit cell volume. Fig.6(b) depicts the variation of equivalent cubic, premartensite (a$_{pm}$ ≈(1/√2) a$_c$, b$_{pm}$ ≈a$_c$, c$_{pm}$ ≈ (1/√2) a$_c$) and martensite (a$_m$ ≈(1/√2) a$_c$, b$_m$ ≈a$_c$, c$_m$ ≈(1/√2) a$_c$) cell parameters and unit cell volume with temperature. It is evident from this figure that unit cell volume of all the three phases decreases with decreasing temperature as expected for normal solids. However, the volume of the premartensite phases drops discontinuously as it transforms to the martensite phase confirming the first order character of this transition. Interestingly no discontinuous change is discernable at the cubic austenite to the premartensite phase transition temperature in Fig 6(b). This is because of the weakly first order nature of this transition. *The observation of hysteresis in the austenite to premartensitic transition temperature during heating and cooling has earlier* *been reported by calorimetric measurements also and has been attributed to the first order nature of this transition [10].*

The adaptive phase model for the formation of the modulated martensite structure of $Ni_2MnGa$ is based on the presumption that the ground state corresponds to the commensurate non- modulated tetragonal $L1_0$ phase. [17, 18] The presence of satellite peaks in the neutron diffraction pattern at 5K clearly [16] rules out this possibility. We revisited the Rietveld refinement carried out by us earlier [16], assuming commensurate modulation for the neutron diffraction pattern recorded at 5K, to check if the ground state of the martensite phase corresponds to a commensurate modulated structure (the so called "lock-in" phase [35]). The value of q obtained using superspace Rietveld refinement at 5K is also found to be an irrational number and corresponds to the value obtained by the extrapolation of the temperature dependence of *q* up to 108K obtained using SXRPD patterns (see Fig 7). Thus the incommensurate martensite does not transform to any commensurate "lock-in" phase.

Incommensurate phase transitions have been investigated both theoretically and experimentally extensively (see e.g. the review by Per Bak [Ref.35])). An incommensurate modulated structure results when *q* is an irrational number. Since there are always rational numbers close to an irrational number, it has been proposed that an incommensurate phase may undergo transition to several intermediate commensurate structures (the "lock-in" phases) on varying some thermodynamic parameter like temperature, pressure, composition, magnetic field etc.[35] Theoretically four different kinds of variations have been predicted for the temperature dependence of *q*: (a) smooth analytic (b) incomplete devil's staircase, (c) complete devil's staircase and (d) harmless staircase behaviour. In (b), (c) and (d), the incommensurate structure may lock into commensurate structures corresponding to different rational approximants. The question arises whether the incommensurate modulated structure of the martensite phase of $Ni_2MnGa$ can lock into commensurate modulated structures at low temperatures either continuously or through a series of rational approximant commensurate structures. The temperature variation of the modulation vector '$q$' shown in Fig. 7 for the cooling cycle reveals smooth analytic behaviour with no evidence for any plateau corresponding to lock-in phases. There is, however, a discontinuous change at the martensite start temperature $M_s^c \sim 218$ K. Similar discontinuous change occurs during heating cycle also. The discontinuous change in $q$ at the premartensite to martensite phase transition temperature further confirms the first order nature of the phase transition.

**Concluding remarks**

We have investigated the austenite-premartensite-martensite phase transition sequence in $Ni_2MnGa$ FSMA using high resolution SXRPD. The premartensite - martensite phase transition is found to be a first order phase transition as revealed by discontinuous change in the unit cell volume and coexistence of the two phases during both cooling and heating cycles. The nanoscale twinning of the $L1_0$ unit cells, postulated in the adaptive phase model, is supposed to achieve an invariant habit plane between the cubic austenite and the non-modulated tetragonal martensite phase. The premartensite phase, where such an invariant habit plane has already been achieved, should not undergo further nanotwinning to give rise to the martensite phase unless both the phases emerge simultaneously in different parts of the austenite crystal. If they emerge simultaneously from the austenite phase, they should continue to coexist in different parts/grains of the same sample as both the phases would have attained the invariant habit plane requirement and continue to coexist at lower temperatures. On the contrary, our results show unambiguously that the premartensite and martensite phases appear successively. This demonstrates that the martensite structure results from the premartensite phase and not from the tetragonal Bain distortion of the austenite phase through repeated nanotwinning to achieve an optimum habit plane. *This rules out the applicability of the adaptive phase model for stoichiometric $Ni_2MnGa$.* Furthermore, the reversibility of the martensite-premartensite transition and the existence of the incommensurate 7M like modulated structure down to 5K without any intermediate lock-in phase suggests that the incommensurate 7M like modulated structure corresponds to the ground state of $Ni_2MnGa$. These results are inexplicable within the framework of the adaptive phase model but put forward the soft phonon mode model as the most plausible model for the origin of modulations in the stoichiometric $Ni_2MnGa$ ferromagnetic shape memory alloy. *However, further investigations are required to test the applicability of the adaptive phase model in off-stoichiometric Ni-Mn-Ga alloy compositions as the present work is confined to the stoichiometric composition only. There have been several first principles calculations [e.g. Ref. 36] showing that the soft phonon is a result of Fermi surface nesting, giving further proof that the modulation results from the soft phonon behavior, and not by the adaptive model.* The observation of charge density wave [37] also reveals the electronic origin of modulation [38]. Last but not the least, the observation of inhomogeneous atomic displacement in the modulated phase [16] and phasons [25] further support our conclusion that adaptive phase model is not applicable to the modulated phase of $Ni_2MnGa$.

**Acknowledgments**

S. S. thanks Alexander von Humboldt foundation, Germany for Research Fellowship. Financial support for the x-ray diffraction studies using synchrotron radiation under the DST-DESY project operated through Saha Institute of Nuclear Physics is gratefully acknowledged. DP thanks the Science and Engineering Research Board of India for the award of J.C. Bose National Fellowship. Parts of this research were carried out at the light source PETRA III at DESY, a member of the Helmholtz Association (HGF), using beamline P02.1.

**Figures:**

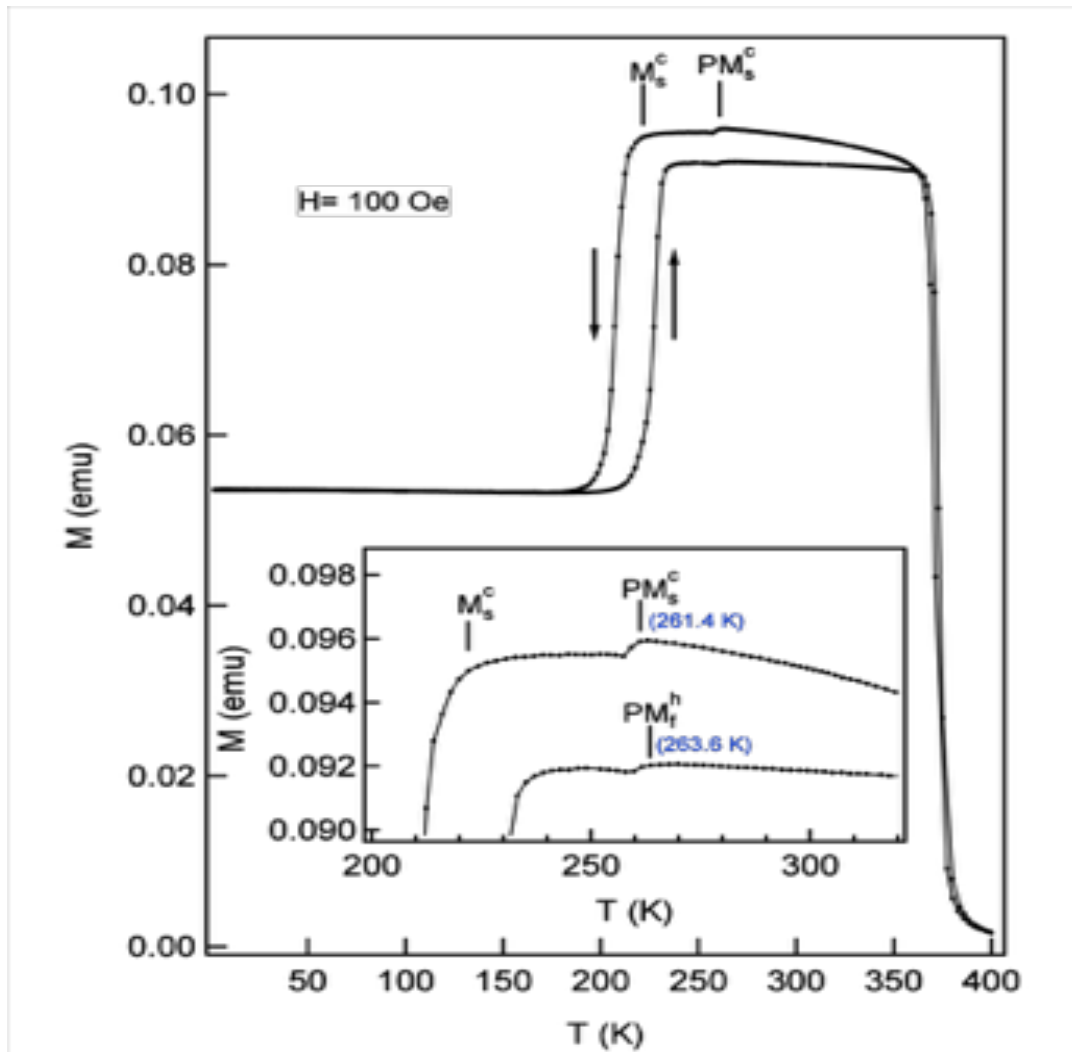

Fig. 1. Low field (at 100 Oe) magnetization as a function of temperature in cooling and heating cycles. Inset shows an expanded view of the austenite to premartensite to martensite (during cooling and martensite to premartensite to austenite (during heating) transitions over 200 to 320 K range.

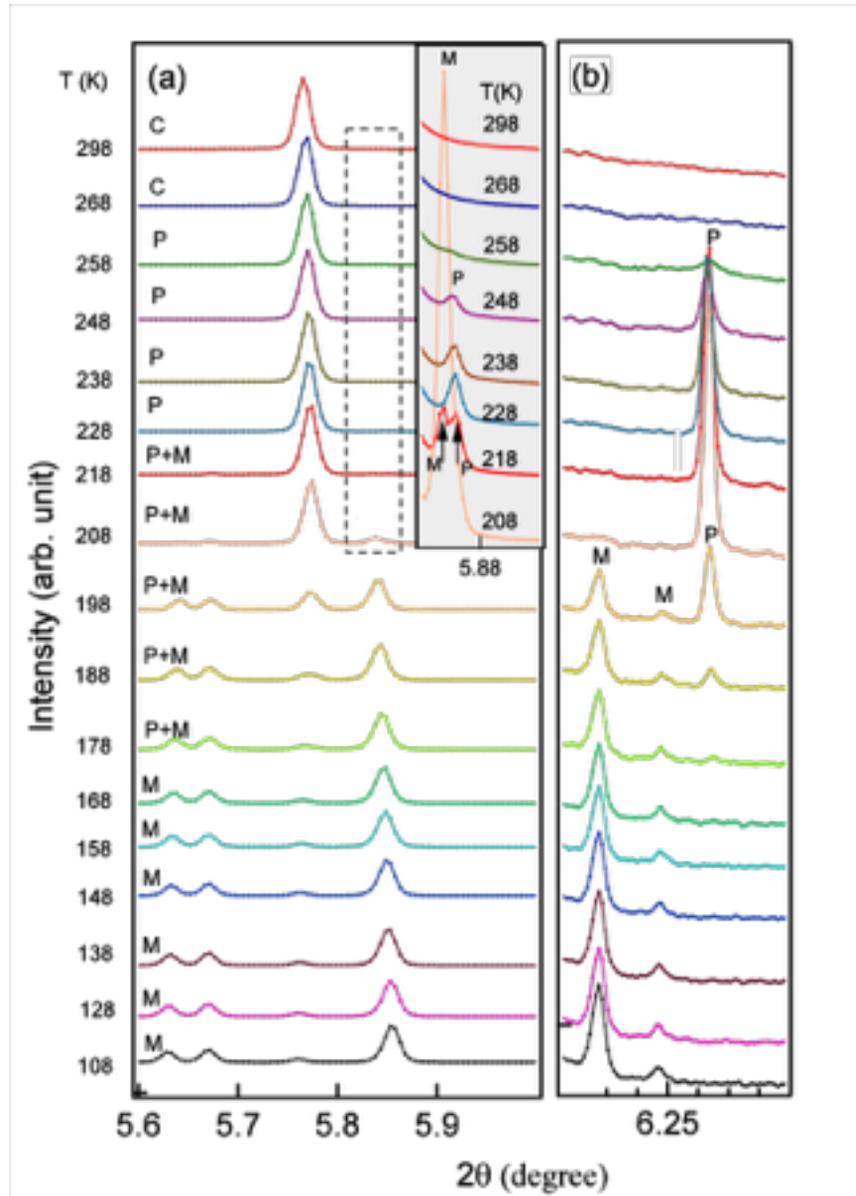

Fig. 2. (color online) Evolution of SXRPD patterns of Ni$_2$MnGa as a function of temperature during cooling cycle. "C", "P" and "M" represent the Bragg peaks due to the cubic, premartensite

and martensite phases, respectively. Inset in (a) shows the Bragg reflections on an expanded scale.

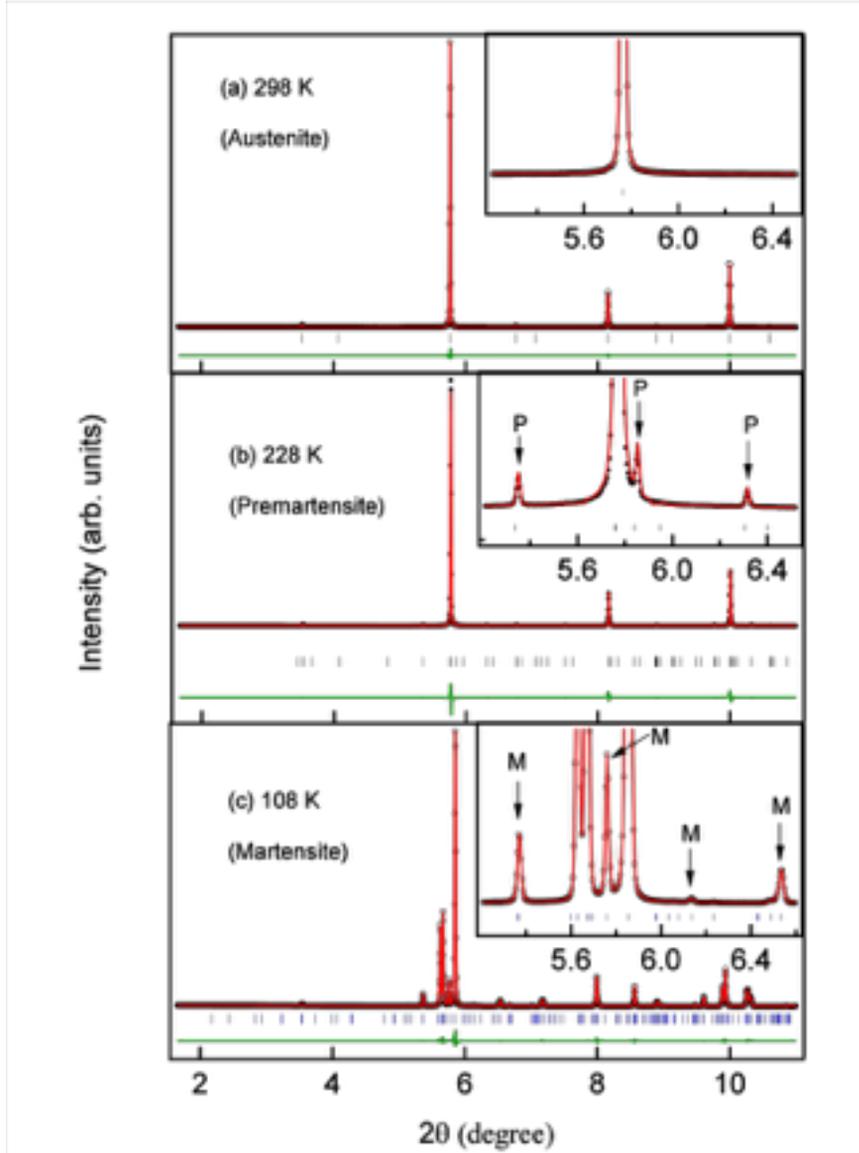

Fig. 3. (color online) Rietveld fits for the SXRPD patterns of $Ni_2MnGa$ at (a) 298 K (cubic austenite phase), (b) 228 K (premartensite phase) and (c) 108 K (martensite phase). The experimental data, fitted curve and the residue are shown by circles (black), continuous line (red) and bottom most plot (green), respectively. The tick marks (blue) represent the Bragg peak

positions. The insets show the fit for the main peak region on an expanded scale. Arrows in (b) and (c) represent satellite reflections in the premartensite (P) and martensite phases (M), respectively.

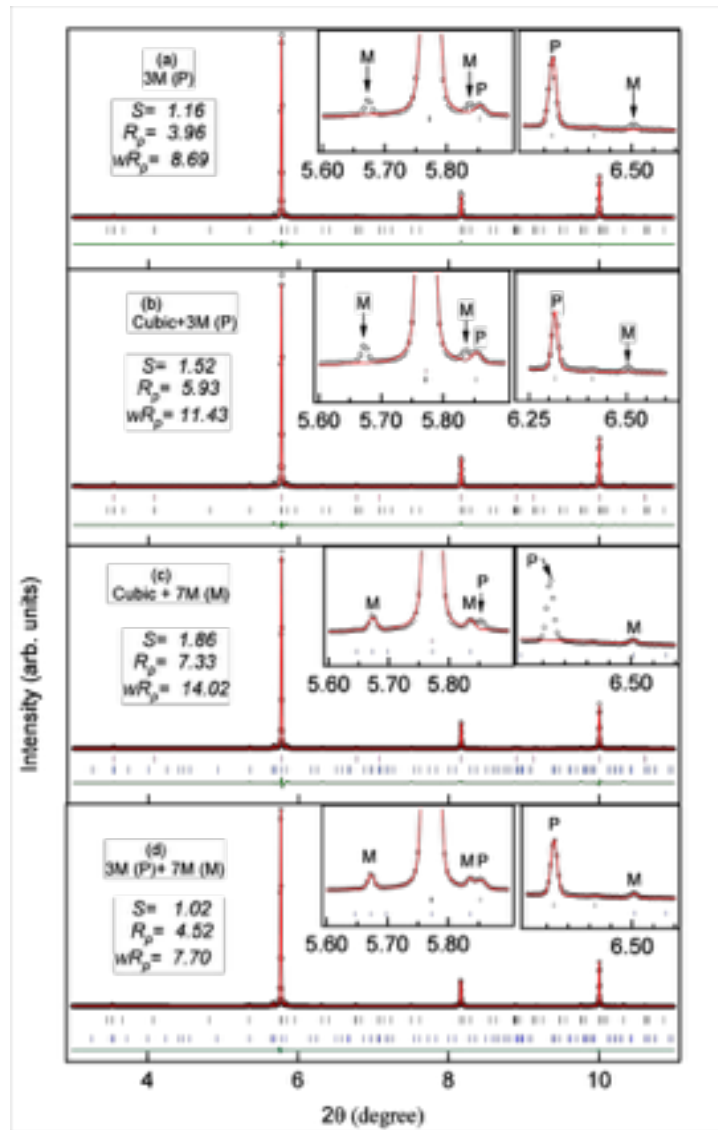

Fig. 4. (color online) Rietveld fits for the SXRPD patterns of Ni$_2$MnGa at 218 K with (a) 3M like incommensurate premartensite structure, (b) combination of cubic and 3M like

incommensurate premartensite structure, (c) combination of cubic and 7M like incommensurate martensite structure and (d) combination of 3M like incommensurate premartensite and 7M like incommensurate martensite structure. The experimental data, fitted curve and the residue are shown by circles (black), continuous line (red) and bottom most plot (green), respectively. The tick marks (blue) represent the Bragg peak positions. The insets show the fit for the satellite reflections corresponding to the premartensite (P) and martensite (M) phases.

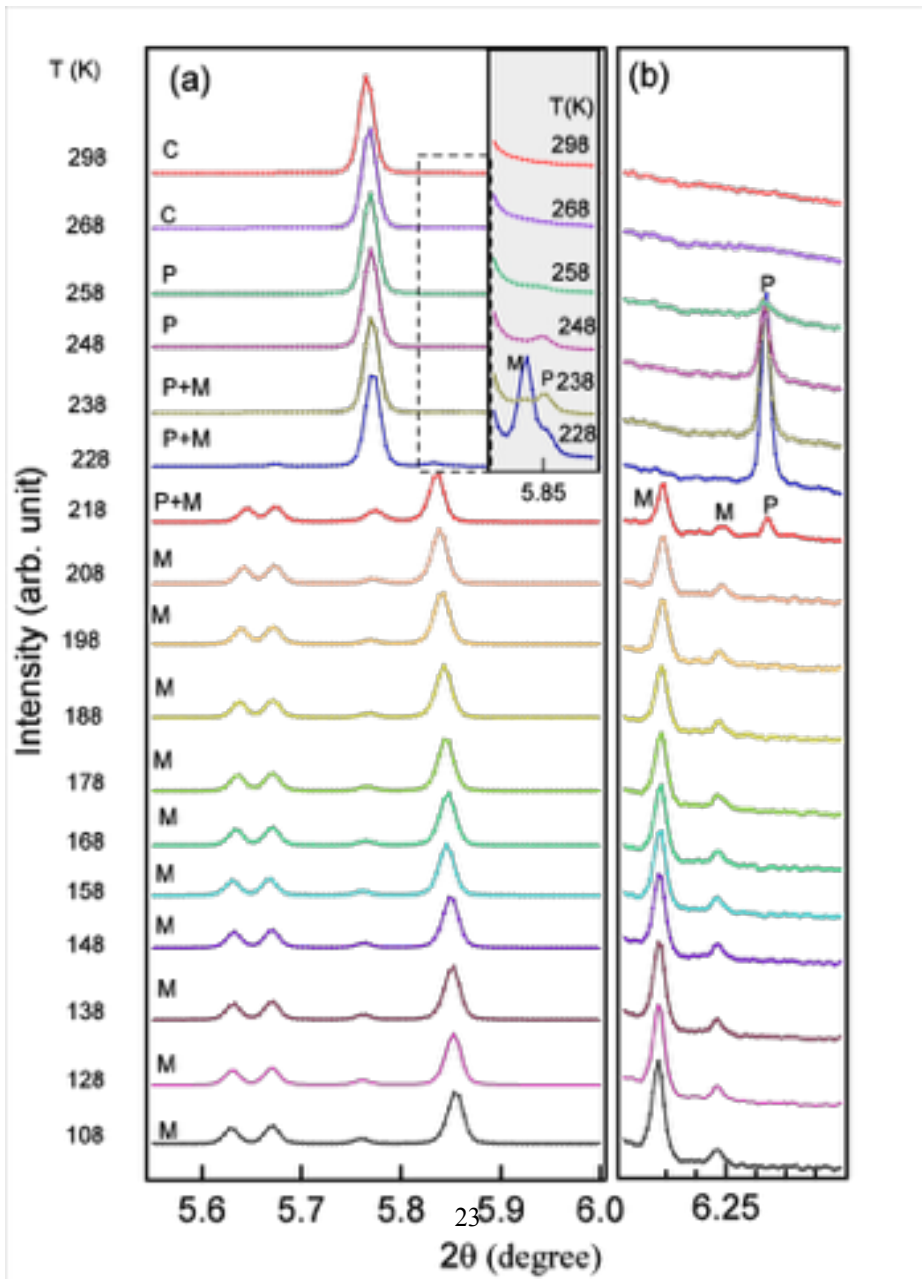

Fig. 5. (color online) Evolution of SXRPD patterns of $Ni_2MnGa$ as a function of temperature during heating cycle. "C" "P" and "M" represent the Bragg peaks due to the cubic, premartensite and martensite phases, respectively. Inset (a) shows the Bragg reflections on an expanded scale.

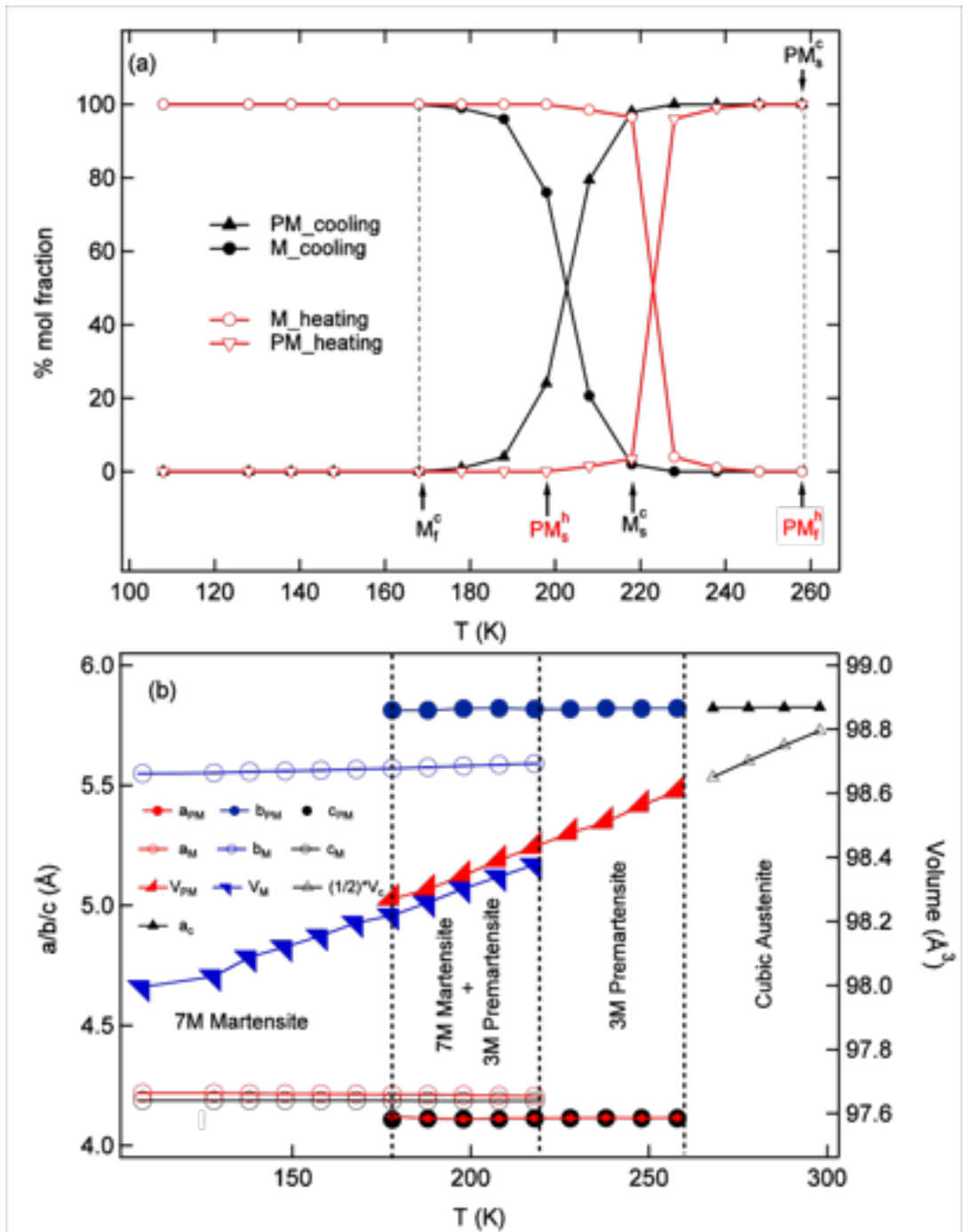

Fig. 6. (color online) (a) Variation of weightfraction of the premartensite (PM) and martensite phases (M), as obtained by Rietveld refinements, with temperature during cooling and heating cycles. (b) Temperature variation of $a$, $b$, $c$, in the austenite (C), 3M like premartensite (PM) and the 7M like martensite (M) phase regions for the cooling cycle. The volume of cubic phase is scaled with ½ for comparison.

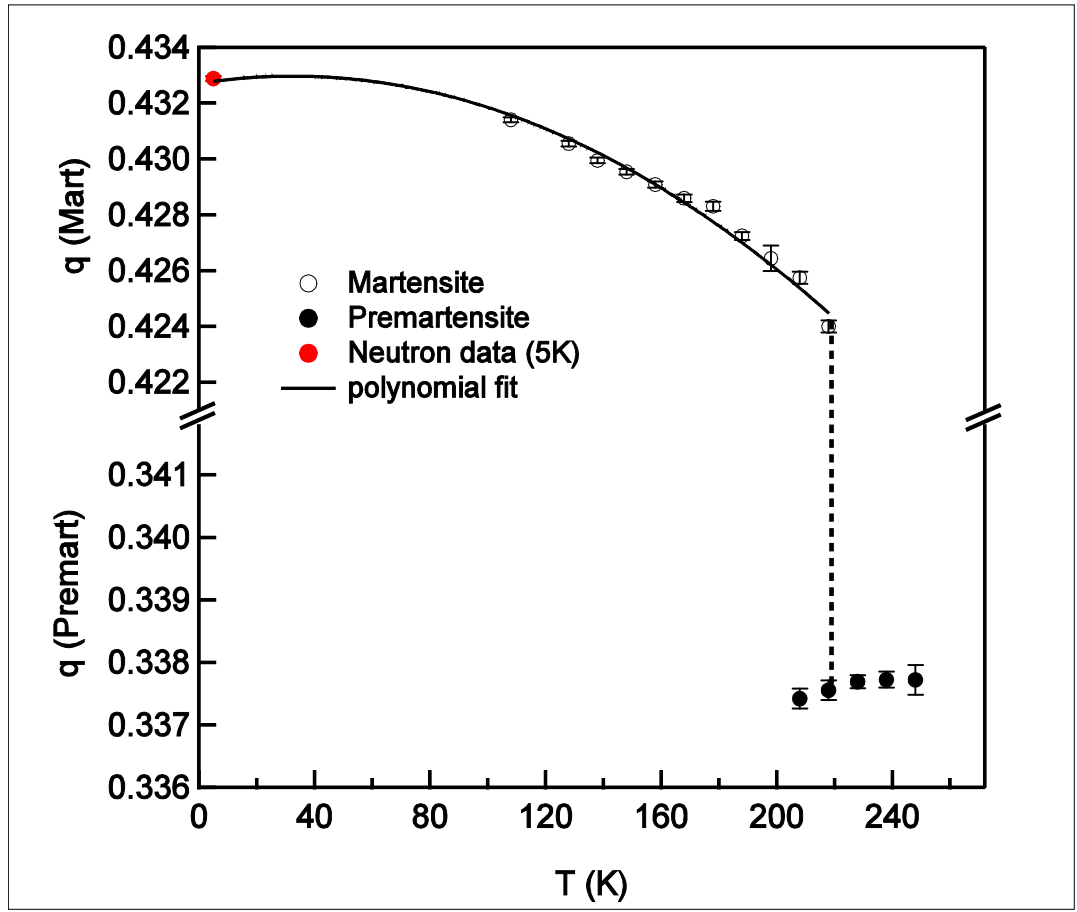

Fig. 7. (color online) Variation of modulation vector (*q*) as a function of temperature obtained from superspace Rietveld refinements during cooling. The modulation wave vector obtained from neutron diffraction data at 5K is also shown in red colour. Dashed line shows discontinous jump in *q* at the transition temperature. *q*-T is nonlinear as *q* behaves like an order parameter that changes discontinuously at a first order phase transition .